\documentclass[journal]{IEEEtran}
\usepackage{color}
\usepackage{graphicx}
\usepackage{multirow}
\usepackage{amsmath}
\usepackage{algorithm}
\usepackage{algorithmic}
\usepackage{subfigure}
\usepackage[table]{xcolor}
\usepackage{array}

\begin{document}
\title{Energy Efficient Sleep Awake Aware (EESAA) Intelligent Sensor Network Routing Protocol}

\author{T. Shah, N. Javaid, T. N. Qureshi\\ \vspace{0.4cm}

        COMSATS Institute of Information Technology,\\ 
        44000, Islamabad, Pakistan. 

        }
\maketitle

\begin{abstract}
Wireless Sensor Networks (WSNs), with growing applications in the environment which are not within human reach have been addressed tremendously in the recent past. For optimized working of network many routing algorithms have been proposed, mainly focusing energy efficiency, network lifetime, clustering processes. Considering homogeneity of network, we proposed Energy Efficient Sleep Awake Aware (EESAA) intelligent routing protocol for WSNs. In our proposed technique we evaluate and enhance certain issues like network stability, network lifetime and cluster head selection process. Utilizing the concept of characteristical pairing among sensor nodes energy utilization is optimized. Simulation results show that our proposed protocolnificantly improved the network parameters and can be a useful approach for WSNs.

\end{abstract}

\begin{IEEEkeywords}
Energy efficient, protocol, sleep, awakw
\end{IEEEkeywords}

\section{Introduction}
Advancement in technologies, devices many opportunities for efficient usage of resources in critical atmospheres. Wireless Sensor Networks (WSNs) brought a revolutionary change in this context. Gathering and delivering of useful information to the destination, became able with advent of this technology. Applications like battlefield surveillance, smart office, traffic monitoring and etc, can be well monitored through such schemes.

WSN is composed of multiple unattended ultra-small, limited-power sensor nodes deployed randomly in the area of interest such as inaccessible ares or disaster places for gathering of useful information. Miniature sensor nodes capable of sensing, processing useful information from, and transmitting to destination has opened many research issues. These battery powered sensor nodes are mounted with limited processing and storage facilities. As WSNs are exposed to dynamic environments, due to such configuration connectivity loss of nodes may degrade the performance of network.

Design of protocols which should be energy efficient and hence, enhancing the network life time is important for better performance. Centralized algorithms are effected badly when a critical node stops working and thus, results in a serious protocol failure. In contrast, distributed protocols can handle such failures more efficiently and can be a suitable solution. Clustering structured routing protocol capable of data aggregation are designed for energy efficiency of a network. Within a cluster localized algorithms can operate without any wait of control messages and hence, reducing the delay. Better scalability is also achieved through these localized algorithms when compared with centralized one's.

In this paper. we evaluate the performance of clustering algorithms on the basis of stability period, network life time and throughput for WSNs. We enhance the above mentioned parameters. Information from sensor nodes is forwarded to cluster heads (CHs) and these CHs are responsible to transmit this information to base station (BS) which is placed far away from the field.

Clustering algorithms like LEACH, and DEEC  [1,5] for sensor networks have achieved reasonable goals regarding better performance of networks. Following their thoughts we proposed a new pairing concept based on applications and specified distances between the sensors which will yield significant improvements in the efficiency of network.

Rest of the paper is organized as follows: Section II describes related work whereas, Section III  describes our protocol EESAA and section IV describes simulations. In the end we concluded the paper.

\section{Related Work}
In WSNs, (Homogenous, Heterogenous) energy always remains a constraint. Many techniques have been proposed to utilize energy of sensor nodes in a better way. Concept of clustering yields significant results in optimizing energy cost for both homogenous and heterogenous networks. In clustering some of the nodes are selected as CHs and had to spent more energy than rest of nodes for a specific period of time. This high energy consumption is due to aggregation and long range transmission of data. Many clustering algorithms e.g., LEACH, PEGASIS, DEEC, SEP, E-SEP [1,2,5,4,6].etc have been proposed which discuses the efficient usage of energy in sensor networks.

CHs in LEACH [1] protocol are selected periodically and energy drains uniformly by role rotation. In PEGASIS [2] energy load is distributed by forming a chain itself or being organized by BS. For such chain formation global knowledge about the network is essential and results in wastage of resources. In DEEC [5] , sensor node are independently elected as CHs based initial energy and residual energy. SEP [4] is designed to deal with heterogenous networks which introduced the concept of advance and normal nodes for cluster head election.

Performance is evaluated on the basis of network stability period, clustering process and throughput. In our EESAA, keeping homogeneity in mind we tried to enhance all these parameters. EESAA keeps the merits of distributed clustering as well.

\section{EESAA: The Proposed Protocol}
In this section, we present a new routing protocol for homogenous networks called EESAA. Our goal is to minimize energy consumption in order to enhance network stability period and network lifetime. For this purpose, we introduced the concept of pairing. Sensor nodes of same application and at minimum distance between them will form a pair for data sensing and communication. In our EESAA protocol, we also enhance CHs selection technique, by selecting CHs on basis of remaining energy of nodes. More comprehensive description of coupling among nodes is defined as follows.

\subsection{Advance Coupling Network Model (ACNM)}
In this section, we explain Advance Coupling Network Model(ACNM). Initially senor nodes measure their location through GPS (Global Positioning System). The nodes transmit their location information, Application type and $Node \_ ID$ to the Base Station (BS). Then, this gathered information is utilized by BS to compute mutual distance between nodes. Nodes which are at minimum distance from each other in their intra cluster transmission range and of same application type are coupled in pair by BS. Then BS broadcast pairing information to all the nodes in network. Nodes become aware of their coupled node. During coupling process some nodes are left out because they are not in inter cluster transmission range of any other node.

According to the proposed scheme, The nodes switch between "Sleep" and "Awake" mode during a single communication interval. Initially node in a pair switch into Awake-mode also called Active-mode if its distance from the BS is less then its coupled node. Node in Active-mode will gather data from surroundings and transmit this data to CHs. During this period transceiver of the coupled node will remain off, and switches into Sleep-mode. Sleep-mode nodes cease their communication with CHs and only sense the network status. In next communication interval,  nodes in Active-mode switch into Sleep-mode and Sleep-mode nodes switch into awake-mode. In this way, we are able to minimize energy consumption because nodes in Sleep-modes save their energy by not communicating with the CHss. Nodes in Sleep-mode also save their energy by avoiding overhearing and idle listening during Sleep-mode. Isolated nodes remain in Active-mode for every round till their energy resources depleted.

\begin{figure*}[t]
\centering
\includegraphics[height=6cm,width=18cm]{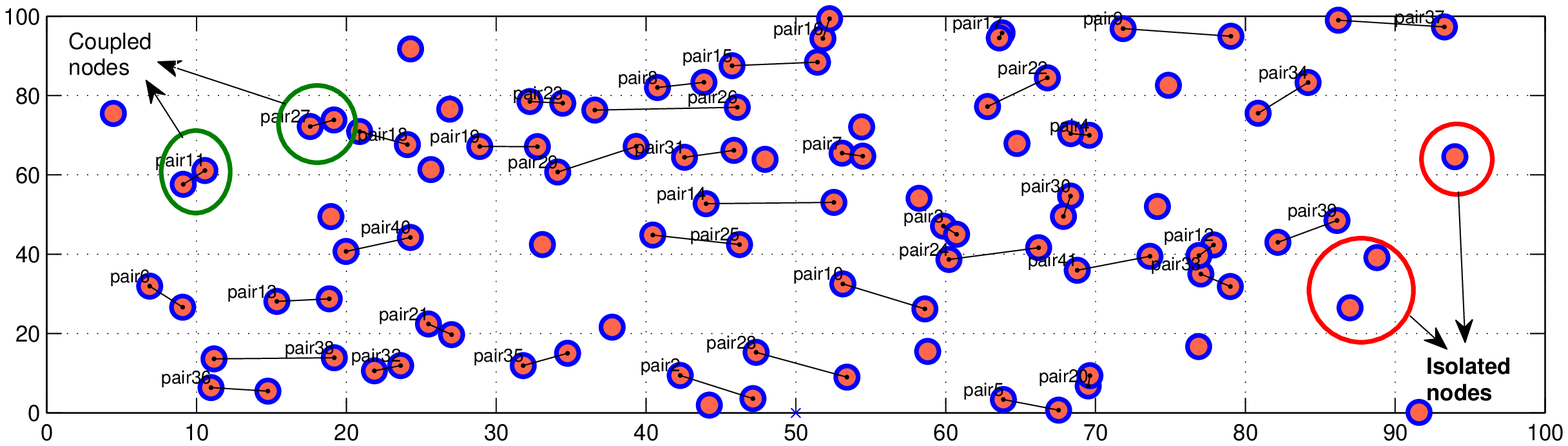}
\caption{Advance Network Coupling Model}\label{Figure 1}
\end{figure*}

\subsection{Network Settling Phase (NSP)}
In NSP, optimal number of CHs are selected with the help of distributed algorithm. Initially all nodes have same energy and network is homogenous in terms of energy level. In LEACH [1] protocol every node decides to become CH or not in current round. The decision is based on desired percentage of CHs per round which is $P_{d}$. In order to assure average number of CH ($P_{d}$$\times N$  ) for N number of nodes, Leach allows each node to become CH after every 1/$P_{d}$ rounds. Number of rounds after which a node become CH refer to as epoch. In homogenous network, energy of nodes after first round can not be same. If the epoch for some high energy nodes and low energy is same, these nodes have same probability to become CHs. There is no proper distribution of CH responsibilities, nodes with low energy will die quickly as compare to nodes with high energy. In our EESAA protocol the CHs selection after first round is based on remaining energy of each node.

Nodes in Active-mode take participation in CH election process. In first round when all nodes have same initial energy $E_{o}$ , nodes in Active-mode will elect them self as CHs on the basis of probability of selecting CH using distributed algorithm. Each node chose a random number between 0 to 1 and compares it to a threshold $T_{h}$, which is calculated as:

\begin{eqnarray}
 T_{h} = \left\{
  \begin{array}{l l}
    \frac {P_{d}}{1 - P_{d}*((firstround)mod\frac{1}{P_{d}})} & \quad \textrm{if $n \in A $ }\\
    0 & \quad \textrm{otherwise}\\
  \end{array} \right.
\end{eqnarray}

where, A is the set of nodes which are in Active-mode in first round. If the random number selected by node is less then threshold $T_{h}$, Node will elect itself as a CH and called as Parent-Cluster-Heads (PCHs). When node has been selected as PCH, it broadcasts an advertisement message to whole network. Only Active-mode nodes hear the broadcast advertisements from different PCHs, They select their PCHs on the basis of Received Signal Strength Indication (RSSI) of advertisements. When an Active-mode node decides to which cluster it wants to associate, it transmits a request to that PCH using Carrier Sensed Multiple Access (CSMA) MAC protocol to avoid collision. Along with request Active-mode nodes also transmit their energy information to the PCH. The PCH computes the remaining energy and its distance from each node and select CH, called Child-Cluster-Head(CCH), for the next round. CCH is selected on the basis of maximum remaining energy of nodes. If different nodes have same remaining energy then a node at minimum distance from PCH is selected as CCH. When PCH selects CCH, it sets up TDMA schedule for associated nodes for communication. PCH then broadcast CCH information and TDMA schedule associated nodes in its cluser. Each node in cluster transmit its data to PCH in its TDMA slot.

\subsection{Network Transmission Phase (NTP)}
In NTP, all nodes in Active-mode, transmit their sensed data to CH during their assigned TDMA slots. Nodes in Sleep-mode do not take participation in NTP and thus save their energy by turning their transceiver off. CHs aggregates received data form each nodes and transmit to BS. Data aggregation is a key signalling technique to compress the amount of data. Due to data aggregation technique a noticeable amount of energy is saved.
If there are N total number of nodes and k are the optimal number of CHs then the average number of nodes in each cluster will be:
\begin{equation}\label{2}
    \left(\frac{N}{K}-1 \right)
\end{equation}

In order to transmit data, The radio of a non-CH node dissipates $E_{TX}$ to run the transmitter circuitry and $E_{amp}$ for transmit amplifier to achieve acceptable SNR (Signal-to-Noise Ratio). So, for transmission of $L_{C}$ bit message a non-CH node expands:
\begin{equation}\label{3}
   E_{non-CH}=\left(\frac{N}{K}-1 \right)(E_{TX}\times L_{C}\times E_{amp}\times L_{C}\times d^{2}_{to_CH}
\end{equation}
To receive data from non-CH node on by the radio of CH in each cluster expands:
\begin{equation}\label{3}
    E_{rec}= ({E}_{RX}\times {L}_{c})\left(\frac{N}{K}-1 \right)
\end{equation}

 where, ${E}_{RX}$ is energy dissipate by receiver circuitry for receiving data. Energy dissipated by CH to aggregate data received from its associated nodes.
\begin{equation}\label{5}
    E_{AGR}= ({E}_{AD} \times {L}_{c})\left(\frac{N}{K} \right)
\end{equation}

 Transmission energy $E_{T}$ dissipates by CH to transmit aggregated data to the BS is:
\begin{equation}\label{6}
    E_{T}=E_{TX} \times L_{A} \times E_{amp} \times L_{A} \times d^{2}_{to BS}
\end{equation}

 where, $L_{A}$ is aggregated data and $d^{2}_{to BS}$ is the distance between CH and BS. Total energy dissipated by CH  a round is:
\begin{equation}\label{7}
    E_{CH}=E_{rec}+E_{AGR}+E_{T}
\end{equation}

 Total energy dissipate by CH is the energy dissipated in reception of data from its associated nodes,aggregation of received data and transmission of that data to the BS.

 \begin{figure*}[t]
\begin{center}
\includegraphics[height=18cm,width=10cm]{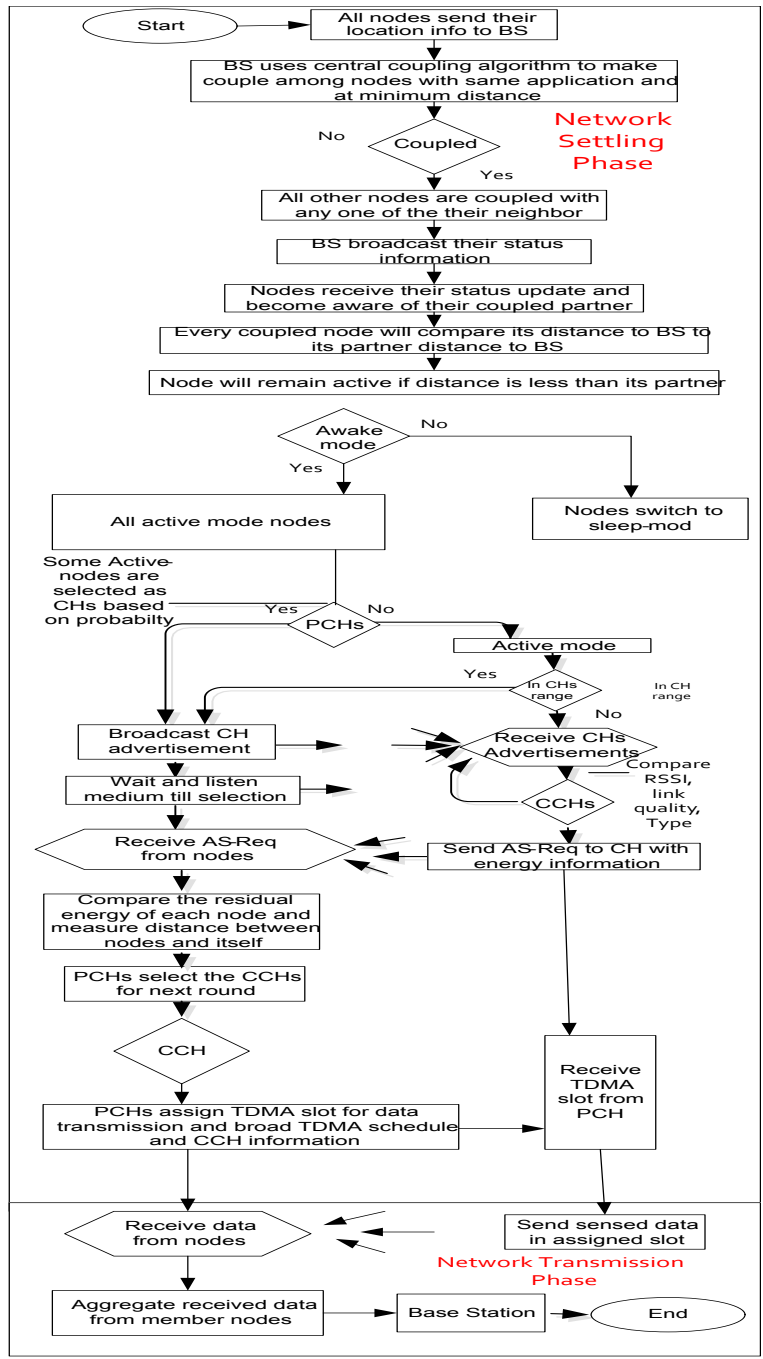}
\caption{Flow chart of EESAA}\label{Figure 33}
\end{center}
\end{figure*}

 \subsection{Node Mode Setup Phase (Node Mode Setup Phase)}
 In this phase, Every node decides whether switch to sleep mode or active mode for the next round. Active-mode node first check that weather it is CCH or not. If it is not CCH it will turn its transceiver off and switch into sleep mode. If it is elected as CCH, it will remain active for the next round. Sleep-mode nodes switch to active if their coupled partner not selected as CCH. Nodes are not coupled with any other node will remain active through its lifetime.

\begin{algorithm}[H]
\caption{: Node Mode Setup Phase}
\begin{algorithmic}[1]
\STATE \textbf{END OF ROUND}
   \IF{( node == coupled )}
     \IF{( node\_ mode==active \&\& CCH\_FLAG==1)}
                \STATE node\_mode=active
         \ELSIF{(node\_mode==active\&\&CCH\_FLAG==0)}
                \STATE node\_mode = sleep
         \ELSIF{(node\_mode==sleep\&\&neighbor\_CCH\_FLAG==1)}
                \STATE node\_mode=sleep
         \ELSIF{(node\_mode==sleep\&\&neighbor\_CCH\_FLAG==0)}
                \STATE node\_mode=active
         \ENDIF
     \ELSIF{(node==coupled\&\&node\_neighbor==dead)}
               \STATE node\_ mode=active
     \ELSE
                \STATE node\_ mode = active
   \ENDIF
\end{algorithmic}
\end{algorithm}

 Above algorithm defines how nodes switch between sleep and awake mode in our EESAA protocol. Node will first check that it is coupled or not. If node is coupled then node check its mode and if it is in Active-mode then it check its CCH flag. If its CCH flag is ON, it will remain in Active-mode. If node is in Active-mode and its CCH flag is not ON it will switch into Sleep-mode. If node is in Sleep-mode, it checks that whether its coupled partner's CCH flag is on or not. If node's coupled partner's CCH flag is on it will remain in sleep mode. If not then node switches to active mode. If coupled partner of a node is dead it will remain active. All that nodes which are left out in coupling process remain active for whole network life time.

 \begin{figure*}[t]
\centering
  \includegraphics[scale=0.5]{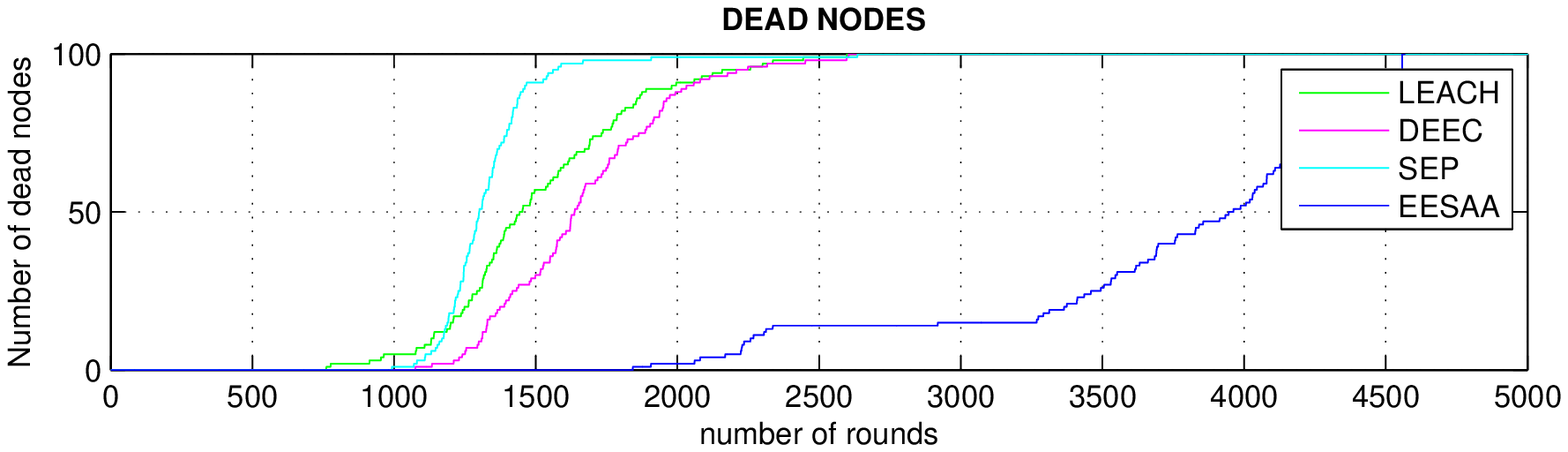}
  \caption{Dead Nodes for $100m\times100m$ Network with 100 nodes}\label{abc}
\end{figure*}

\begin{figure*}[t]
\centering
  \includegraphics[scale=0.5]{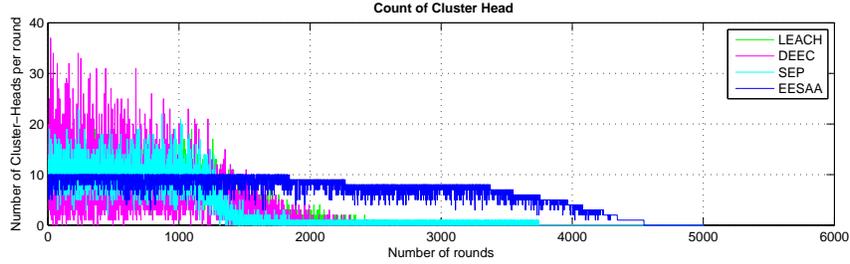}
  \caption{CHs per round}\label{abc}
\end{figure*}

\begin{figure*}[t]
\centering
  \includegraphics[scale=0.5]{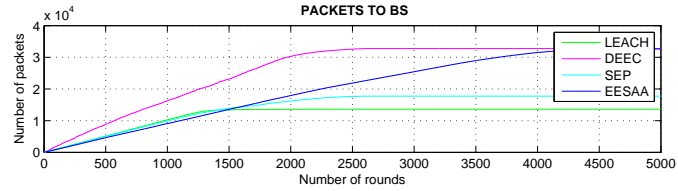}
  \caption{Packet to BS Nodes for $100m\times100m$ Network with 100 nodes}\label{abc}
\end{figure*}

 \section{Simulation Results}
 We measure the performance of our proposed protocol EESAA by performing comparative simulations. In our simulations we generate a sensor field of $100m\times100m$ size. In this field we randomly drop $n \epsilon (100)$ sensor nodes with initially energy Eo. Parameters for our simulation are given in table I.

\begin{table}[h]
  \centering
  \caption{Simulation Parameters}
    \begin{tabular}{rrrrrrrrrr|rr}
    %\toprule
    Parameter                                                           &       &       &       & value &  \\ \hline
    %\midrule
    Network size        &       &       &       &            100m * 100m      &  \\\hline
    Initial Energy        &       &       &       &                .5 J      &  \\\hline
    $P_{d} $      &       &       &       &                      .1 J      &  \\\hline
    Data Aggregation Energy cost      &       &       &       &                      50pj/bit j      &  \\\hline
    Number of nodes       &       &       &       &                     100      &  \\\hline
    Packet size        &       &       &       &             4000 bit       &  \\\hline
    Transmitter Electronics (EelectTx) &       &       &       & 50 nJ/bit      &  \\\hline
    Receiver Electronics (EelecRx)     &       &       &       & 50 nJ/bit &  \\\hline
    Transmit amplifier (Eamp)            &       &       &       & 100 pJ/bit/m2 &  \\\hline

    %\bottomrule
    \end{tabular}%
\end{table}%

For analyzing and comparing the performance of EESAA protocol with LEACH, SEP and DEEC protocols we consider the following metrics as given in [1,3,4].

\begin{enumerate}
  \item Stability period: It is duration of network operation from start till first node dies.
  \item Network lifetime: Network lifetime is duration from start till last node is alive.
   \item Instability period: It is duration of network operation from first node dies till the least node dies.
  \item Number of Cluster-heads:It indicates the number of clusters generated per round.
  \item Packet to BS: It is rate of successful data delivery to BS from CHs.
  \end{enumerate}

\begin{figure*}[t]
\centering
  \includegraphics[scale=0.5]{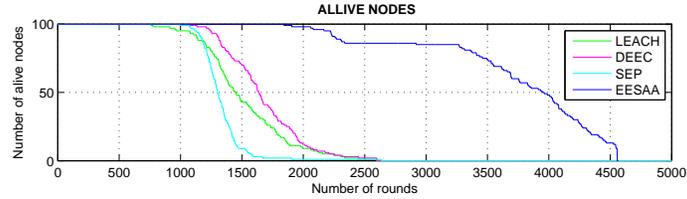}
  \caption{Alive Nodes for $100m\times100m$ Network with 100 nodes}\label{abc}
\end{figure*}

We analyze network lifetime of LEACH, SEP, DEEC and our EESAA protocols. We examine the way the number of alive nodes varies as network evolves. In fig 3 it is depicted that EESAA has a prolong stability period as compare to the other protocols. In EESAA first node dies around 1800 round. Stability period of EESAA is almost 120$\%$ 50$\%$ and 35$\%$ greater than LEACH, SEP and DEEC respectively. From fig 3 it is also depicted that EESAA has 100$\%$ 102$\%$ and 50$\%$ maximum network life time as compared to LEACH, SEP and DEEC. It is because of sleep-awake property of nodes and effective cluster-head selection's algorithm.

From Fig.4 we notice that the first node dies around $1700$ and last node dies after $4000$ rounds. This shows that in EESAA instable region starts very later as compare to other protocols. Figure 5 also shows that there is sudden increase in number of dead nodes in SEP and LEACH protocols whilst in EESAA nodes dies at a constant rate. This observation depicts that in EESAA energy dissipation is properly distributed among all the nodes in the network which in result increases network lifetime.

In Fig.5 we analyze the number of CHs selected in every round for all routing protocols. As shown in Fig.6 SEP, DEEC, LEACH has more uncertainties in CHs selection. Random number of CHs are selected in every round but ESSA has some patterns and controlled CHs selection. EESAA efficient CHs selection algorithm helps it in better and constant data rate transmission to BS. Although EESAA has sleep-awake policy for nodes and less number of data is transmitted to BS however, ESSAA successful data delivery to BS is much better than SEP and LEACH although SEP and LEACH is transmitting data continuously. Other main reason of higher data rate achievement is longer network life time of EESAA. Successful data delivery is shown in Fig 6.

\section{Conclusion}
In this paper, we presented a more optimized routing scheme for WSNs. Main focus was to enhance cluster-head selection process. In EESAA, CHs ale selected on the basis of remaining energy. In EESAA nodes also switches between sleep and active modes in order to minimize energy consumption. In our proposed strategy, stability period of network, and life time has been optimized. Simulation results show that their is significant improvement in all these parameters when compared with some of the existing routing protocols e.g., SEP, LEACH and DEEC.

%\begin{itemize}
%\item Topmost list
%\end{itemize}
%\begin{itemize}
%\item Second level
%\begin{itemize}
%\item Third level
%
%\begin{itemize}
%\item Fourth
%\end{itemize}\end{itemize}\end{itemize}

%
\end{document}